\documentclass[aps,pra,twocolumn,groupedaddress,floatfix]{revtex4}
\usepackage{bm} \usepackage{graphicx}
\usepackage{epsfig}
\usepackage{subeqn}

\begin{document}


\title{Dipolar Bose-Einstein condensates at Finite temperature}

\author{Shai Ronen} 
\affiliation{JILA and Department of Physics, University of Colorado, Boulder, CO 80309-0440}
\author{John L. Bohn} 
\affiliation{JILA, NIST, and Department of Physics, University of Colorado,
 Boulder, CO 80309-0440 }
\email{bohn@murphy.colorado.edu}

\affiliation{}


\date{\today}

\begin{abstract}

We study a Bose-Einstein condensate (BEC) of a dilute gas with dipolar
interactions, at finite temperature, using the Hartree-Fock-Bogoliubov
(HFB) theory within the Popov approximation. An additional
approximation involving the dipolar exchange interaction is made to
facilitate the computation. We calculate the temperature dependence of
the condensate fraction of a condensate confined in a cylindrically
symmetric harmonic trap. We show that the bi-concave shaped
condensates found in Ref.~\cite{Ronen07} in certain pancake traps at
zero temperature, are also stable at finite temperature. Surprisingly,
the dip in the central density of these structured condensates is
actually enhanced at low finite temperatures. We explain this effect.

\end{abstract}


\maketitle

\section{Introduction \label{intro}}

The realization of a Bose-Einstein condensate (BEC) of $^{52}$Cr
\cite{Stuhler05} marked a major development in degenerate quantum gases
in that the inter-particle interaction via magnetic dipoles in this
BEC is much larger than that in alkali atoms, and leads to an
observable change in the shape of the condensate. The long range
nature and anisotropy of the dipolar interaction pose challenging
questions about the stability of the BEC and have led to predictions of  unique
phenomena, such as roton-maxon spectrum, different phases of vortex
lattices, and bi-concave shaped condensates
\cite{Yi00,Goral00,Baranov02,Goral02,Santos03,ODell04,Nho05,Cooper05,Pu06,Ronen06a,Ronen07}.

The effect of finite temperature on these phenomena, or new
temperature-dependent effects, remain largely unexplored. Theoretical
finite temperature studies have been confined to path integral Monte
Carlo simulations \cite{Nho05} of a small number (~100) of particles,
and to a homogeneous, quasi 1D system \cite{Mazets04}. In the latter
work, the Popov approximation to the HFB theory has been applied
\cite{Griffin96,Hutchinson97}.  In systems with short range
interaction, the Popov approximation has been found to give excitation
spectra in good agreement with experiment for temperatures up to half
the critical temperature for condensation \cite{Dodd98}. For the
density profile, good accuracy was shown for even higher temperatures,
up to the critical temperature.

In \cite{Ronen06a} we have introduced a new computational algorithm
which allowed us to calculate the Bogoliubov excitation spectrum of
dipolar condensates in cylindrically symmetric 3D traps, at zero
temperature. Here we extend this work in a natural way to describe
finite temperature properties in the Popov approximation. In essence,
the Popov approximation is a self-consistent solution in which the
Bogoliubov excitation spectrum is computed and the different
excitation modes populated according to Bose statistics. This leads to
depletion of the condensate, and thus a shift in excitation
frequencies. The excitation spectrum is re-calculated iteratively
until self-consistency is achieved.

\section{Formalism}

The HFB-Popov equations for the case of short range interactions have
been described by Griffin\cite{Griffin96}. The generalization to long
range interactions is straightforward, and we therefore briefly
formulate it in this section. The confined Bose gas is portrayed as a
thermodynamic equilibrium system under the grand canonical ensemble
whose thermodynamic variables are the temperature $T$ and the chemical
potential $\mu$.  There is a one-to-one relationship between the
chemical potential $\mu$ and the total number of particles $N$.  Below
the critical temperature, $N_0$ of the atoms are in a condensate
state. The system Hamiltonian $K$ for the system with a fixed chemical
potential is then obtained from the Hamiltonian $H$ of the system with
fixed number of total particles via a Legendre transform, and has the
form
\begin{eqnarray}
K=H-\mu N=\int d\bm{r} \hat{\Psi}^{\dagger}(\bm{r})(H_0-\mu)\hat{\Psi}(\bm{r})
\nonumber \\
+\frac{1}{2} \int \hat{\Psi}^{\dagger}(\bm{r}) \hat{\Psi}^{\dagger}(\bm{r'}) 
V(\bm{r'}-\bm{r})\hat{\Psi}(\bm{r'})\hat{\Psi}(\bm{r}), 
\label{eqn:K}
\end{eqnarray}
where $\hat{\Psi}(\bm{r})$ is the Bose field operator that annihilates an atom at position
$\bm{r}$; $H_0=(-\hbar^2/2M)\nabla^2+V_{trap}(\bm{r})$ contains the kinetic energy and the trap potential, and $V(\bm{r})$ is the
particle-particle interaction potential. For the system treated here, the trap potential
is cylindrically symmetric: $V_{trap}(\bm{r})=M(\omega_{\rho}^2\rho^2+\omega_z^2z^2)/2$, where $M$ is the atomic mass, and
$\omega_{\rho}$ and $\omega_z$ are the radial and axial trap frequencies, respectively. 
For polar gases the potential $V(\bm{r})$ may be written:
\begin{eqnarray}
V(\bm{r})=\frac{4\pi\hbar^{2}a}{M}\delta(\bm{r}) + d^2\frac{1-3\cos^{2}\theta}{r^3},
\label{pseudo}
\end{eqnarray}
where $a$ is the scattering length, $d$ the dipole moment, $\bm{r}$
the distance between the dipoles, and $\theta$ is the angle between the
vector $\bm{r}$ and the direction of polarization, which is aligned
along the trap $z$-axis .


The Bose field operator is decomposed into a $c$-number condensate
wave function plus an operator describing the non-condensate part,
$\hat{\Psi}(\bm{r})= \sqrt{N_0}\phi(\bm{r})+\tilde{\psi}(\bm{r})$, and
inserted into Eq.~(\ref{eqn:K}). Terms cubic and quartic in
$\tilde{\psi}(\bm{r})$ are treated within the mean-field approximation
and the grand-canonical Hamiltonian reduces to a sum of three terms:
$K=K_0+K_1+K_2$. the first term $K_0$ is a $c$ number, the second
term, $K_1$, is linear in $\tilde{\psi}(\bm{r})$ (and its hermitian
conjugate), and the last term, $K_2$, is quadratic in these
quantities. Within the Popov approximation, the so called anomalous
terms arising from mean field averages of the form
$<\tilde{\psi}\tilde{\psi}>$ are ignored and only  'normal' terms
of the form $<\tilde{\psi}^{\dagger}\tilde{\psi}>$ are included
\cite{Griffin96}.  For a system in equilibrium the linear term $K_1$
is required to vanish identically, giving a generalized
Gross-Pitaevskii (GP) equation for $\phi(\bm{r})$:
 
\begin{eqnarray}
\left[H_0+\int d\bm{r'} (N_0|\phi(\bm{r'})|^2+\tilde{n}(\bm{r'}) V(\bm{r'}-\bm{r})\right]\phi(\bm{r}) \nonumber \\ 
+\int \tilde{n}(\bm{r'},\bm{r})V(\bm{r'}-\bm{r})\phi(\bm{r'})=\mu\phi(\bm{r}),
\label{eq:gpe}
\end{eqnarray}
where $\tilde{n}(\bm{r})\equiv \langle
\tilde{\psi}^{\dagger}(\bm{r})\tilde{\psi}(\bm{r}) \rangle$ is the
density of the non-condensate (thermal) atoms, and $\tilde{n}(\bm{r'},\bm{r})
\equiv \langle \tilde{\psi}^{\dagger}(\bm{r'})\tilde{\psi}(\bm{r}) \rangle$ 
is the 1-particle reduced density matrix, or the correlation
function, of the non-condensate atoms. The term involving
$\tilde{n}(\bm{r})$ represents the mean field contribution due to
direct interaction between thermal cloud and the condensate. The term
involving $\tilde{n}(\bm{r'},\bm{r})$ represents the contribution of
exchange interaction between the thermal cloud and the condensate.
Note that for $V$ with only short range interaction, the exchange term
reduces in form to that of the direct one. But for long range
interaction, the non-local correlation function is needed.  Also, note
that a more careful treatment \cite{Leggett01,Esry97} reveals that,
for an atom number conserving system, the factor $N_0$ in Eq.~(\ref{eq:gpe})
should be replaced by $N_0-1$. This correction is negligible for
$N_0>>1$. 

The term $K_2$ has the form
\begin{eqnarray}
K_2&=&\int d\bm{r} \tilde{\psi}^{\dagger}(\bm{r}){\cal L} \tilde{\psi}(\bm{r}) \nonumber \\
&+&\int d\bm{r}d\bm{r'}\tilde {\psi}^{\dagger}(\bm{r'})n(\bm{r},\bm{r'})V(\bm{r'}-\bm{r})\tilde{\psi}(\bm{r}) \nonumber \\
&+&\frac{N_0}{2}\int d\bm{r} d\bm{r'} \tilde{\psi}^{\dagger}(\bm{r'})
\tilde{\psi}^{\dagger}(\bm{r})V(\bm{r'}-\bm{r})\phi(\bm{r})\phi(\bm{r'}) \nonumber \\
&+&\frac{N_0}{2}\int d\bm{r} d\bm{r'} \tilde{\psi}(\bm{r'})
\tilde{\psi}(\bm{r})V(\bm{r'}-\bm{r}) \phi^{*}(\bm{r})\phi^{*}(\bm{r'}),
\end{eqnarray}
where ${\cal L}=H_0-\mu+\int d\bm{r'}V(\bm{r'}-\bm{r})n(\bm{r'})$, in which 
$n(\bm{r'})=\tilde{n}(\bm{r'})+N_0|\phi(\bm{r'})|^2$ is the total density,
and $n(\bm{r},\bm{r'})=\tilde{n}(\bm{r},\bm{r'})+N_0\phi^{*}(\bm{r})\phi(\bm{r'})$ is the
total correlation function.

The term $K_2$ can be diagonalized by the Bogoliubov transformation,
\begin{eqnarray}
\tilde{\psi}(\bm{r})=\sum_{j}[u_j(\bm{r})\alpha_j+\upsilon^*_j(\bm{r})\alpha_j^{\dagger}], 
\label{eq:transform}
\end{eqnarray}
if the quasi-particle amplitudes $u_j(\bm{r})$ and $\upsilon_j(\bm{r})$ satisfy the coupled HFB-Popov
equations:
\begin{subequations}
\begin{eqnarray}
E_j u_j(\bm{r})&=& {\cal L}u_j(\bm{r}) \nonumber+\int d\bm{r'}  V(\bm{r'}-\bm{r})n(\bm{r'},\bm{r}) u_j(\bm{r'})  \nonumber \\
&+& N_0 \int d\bm{r'} \phi(\bm{r'})V(\bm{r'}-\bm{r})\upsilon(\bm{r'}) \phi(\bm{r}), 
\end{eqnarray}
\begin{eqnarray}
E_j\upsilon_j(\bm{r})&=&{\cal L}\upsilon_j(\bm{r})+\int d\bm{r'} V(\bm{r'}-\bm{r})n(\bm{r},\bm{r'})\upsilon_j(\bm{r'}) \nonumber \\
&+&N_0 \int d\bm{r'} \phi(\bm{r'})^{*}V(\bm{r'}-\bm{r}) u(\bm{r'})\phi(\bm{r})^{*}. 
\end{eqnarray}
\label{eq:popov}
\end{subequations}
The quasi-particle annihilation and creation operators $\alpha_j$ and
$\alpha_j^{\dagger}$ satisfy the usual Bose commutation relations.

In terms of $u$'s and  $\upsilon$'s, the thermal density correlation function is written as:
\begin{eqnarray}
\tilde{n}(\bm{r'},\bm{r})=\sum_j \left[ u_j^*(\bm{r'})u_j(\bm{r})+\upsilon_j(\bm{r'})\upsilon_j^*(\bm{r}) \right] N_{ex}(E_j) \nonumber \\
+\upsilon_j(\bm{r'})\upsilon_j^*(\bm{r}),
\label{eq:denuv}
\end{eqnarray}
where $N_{ex}(E_j)$ is the Bose distribution for the quasi-particle excitations:
\begin{eqnarray}
N_{ex}(E_j) \equiv \langle \hat{\alpha}_j^{\dag}\hat{\alpha}_j\rangle=\frac{1}{\exp\left(\frac{E_j}{k_B T}\right)-1}.
\end{eqnarray} 
The expression for $\tilde{n}(\bm{r})$ in terms of
$u,\upsilon$ is obtained by setting $\bm{r'}=\bm{r}$ in
Eq.~(\ref{eq:denuv}). Similar expressions may be easily obtained for the
total and release energy of the condensate in the trap.

As usual, the self-consistent HFB-Popov equations (\ref{eq:gpe}) and (\ref{eq:popov}) are solved iteratively.
At first, one sets $\tilde{n}(\bm{r'},\bm{r})=\tilde{n}(\bm{r})=0$. The thermal component contribution is then updated at each further step
using Eq.~(\ref{eq:denuv}) until convergence is reached. 

It can be appreciated that long range interactions present a
significant challenge to the computational implementation of the
HFB-Popov method. The difficulties arise even for zero temperature,
where (ignoring the negligible quantum depletion) the HFB-Popov
equations reduce to the Bogoliubov-De Gennes (BdG) equations, which
due to the long range exchange interaction, are now
integro-differential rather than simply differential
equations. In Ref.~\cite{Ronen06a} we have introduced a new algorithm,
which enabled us to solve the BdG equations for a gas with dipolar
interactions in a 3D trap with cylindrical symmetry, by utilizing the
cylindrical symmetry to reduce the effective dimensionality of the
problem from 3D to 2D. For the finite temperatures with which the
HFB-Popov method is concerned, the long range interactions introduce
an additional difficulty: Eqs.~(\ref{eq:gpe}) and (\ref{eq:popov})
involve not only the thermal density $\tilde{n}(\bm{r})$, but also the
thermal correlation function $\tilde{n}(\bm{r'},\bm{r})$. The most
difficult terms are those involving the long range exchange
interaction such as $\int d\bm{r'} V(\bm{r'}-\bm{r})n(\bm{r'},\bm{r})
u_j(\bm{r'})$ in Eq.~(\ref{eq:popov}). In the case where there is no
thermal component, the total correlation function is expressed as a
direct product: $n(\bm{r'},\bm{r})=\phi^*(\bm{r'})\phi(\bm{r})$. In
this case, the above exchange term may be effectively evaluated by the
use of Hankel-Fourier transform \cite{Ronen06a}.
In the presence of a thermal component, the total correlation function does
not have such a direct product decomposition. 
It might be possible to circumvent this complication by direct
evaluation of the exchange integral on a spatial grid (without resort
to Fourier transforms; see for example \cite{Yi01}) - however, this is
complicated by the singular nature ($1/\bm{r}^3$ behavior) of dipolar
interactions at the origin.

To obtain a feasible numerically solvable problem, we therefore make
an additional approximation: in Eqs.~(\ref{eq:gpe}) and
(\ref{eq:popov}) we let $\tilde{n}(\bm{r'},\bm{r})=0$ for $\bm{r'}\neq
\bm{r}$, or equivalently, we let
$n(\bm{r'},\bm{r})=\phi(\bm{r'})^*\phi(\bm{r})$. Physically, in
Eq.~(\ref{eq:gpe}) this amounts to ignoring the forces on the
condensed part due to the long-range exchange interaction with the
thermalized part. In Eq.~(\ref{eq:popov}) this amounts to ignoring
the long range exchange interaction between the thermal component of
the gas and itself. On the other hand, in Eq.~(\ref{eq:popov}) we
do take into account the effect on excitation modes and
frequencies due to the the long range exchange interaction with the
condensate.  We treat exactly the long
range direct interactions which involve $n(\bm{r})$. We also treat exactly the short range (contact) interaction,
for which the exchange and direct terms are identical.

A partial justification for the above scheme may be found in the good
agreement between the ``two-gas model'' of dilute BECs and the full
HFB-Popov description for gases with short range interactions
\cite{Dodd99}. In the two gas description, the condensate wavefunction
is that of a $T=0$ BEC with the appropriate (depleted) number of
condensate atoms $N_0$, and the surrounding thermal cloud is described
by the statistical mechanics of an ideal gas in the combined
potentials of the trap and the cloud-condensate interaction. The full
HFB-Popov description may be reduced to the two gas description by
letting $\tilde{n}(\bm{r,r'})=0$ for all $\bm{r},\bm{r'}$, so that, in
particular, $\tilde{n}(\bm{r}) \equiv \tilde{n}(\bm{r},\bm{r})=0$.
The reason for the success of the two-gas model seems to derive from
the fact that the thermal component is typically much more dilute than
the condensate part.  Therefore, to a good approximation, it may be
described as an ideal gas. The approach suggested above for the
treatment of dipolar BEC at $T>0$ may be described as treating the
thermal component as 'partly ideal', i.e, 'ideal' only with respect to
long range exchange interactions. Thus, it is a compromise between the
full HFB-Popov method and the two gas description.

Moreover, We note as a general thermodynamical property that the
correlation function $\tilde{n}(\bm{r'},\bm{r})$ naturally decreases
towards zero with increase in temperature (for $\bm{r'}\neq\bm{r}$).  Thus,
with increasing temperature, it makes sense to ignore the thermal long
range exchange interaction which is due to the correlation function of
the thermal component of the gas. 

The number of excitation modes that need to be taken into account in
Eq.~(\ref{eq:popov}) in order to saturate the thermal cloud density
is very large (tens of thousands). For the higher excitation modes, the semi-classical
description has proved very useful and accurate\cite{Reidl99,Ronen06a}. Thus, we
follow the approach of solving Eqs.~(\ref{eq:popov}) for the discrete
modes up to an appropriate energy cutoff, and using their semi-classical
version \cite{Ronen06a} for modes above this energy cutoff. The energy cutoff is typically somewhat larger
than the chemical potential and is adjusted in each specific case until convergence is achieved.

\section{Results \label{sec:results}}

\subsection {Cr in a pancake trap}

We first study the effects of temperature for a $^{52}$Cr gas in a
trap with frequencies $\omega_{\rho}=2\pi \times 100 Hz$ and
$\omega_z=2\pi \times 400 Hz$ (a pancake trap).  The magnetic dipole
moment of polarized Cr is relatively large for atoms, 6 Bohr
magnetons. However, the resulting dipole-dipole interaction is still
small compared to the strength of the short range interaction
(scattering length $a=96 a_0$ \cite{Griesmaier06a}). A useful
parameter here is the dimensionless quantity $\epsilon_{dd}=\frac{ m
d^2}{3 \hbar^2/a}$. A homogeneous condensate is unstable if
$\epsilon_{dd}>1$ \cite{Eberlein05}. For Cr, $\epsilon_{dd}=0.16$. However, using
a Feshbach resonance, it is possible to reduce the scattering length and
thus increase the dipolar effects (T. Pfau, private
communication). For the present study we assume a reduced scattering
length of $20 a_0$, so that $\epsilon_{dd}=0.8$.  The number of atoms
in the trap is taken to be $10^5$.

\begin{figure}
\resizebox{3.5in}{!}{\includegraphics{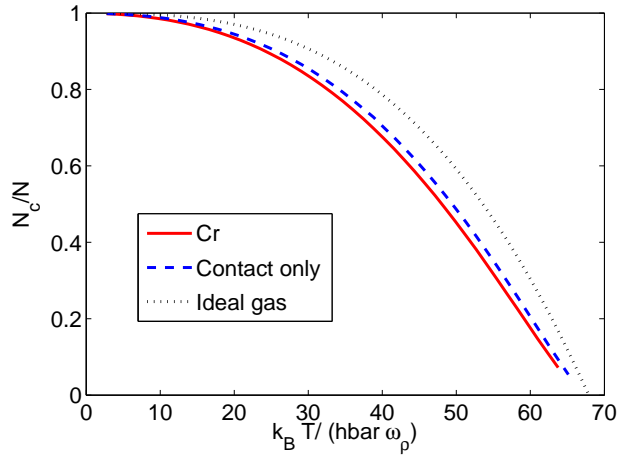}}
\caption {Condensate fraction as a function of temperature
in a pancake trap with aspect ratio 1:4  (i.e, $\omega_{\rho}/ \omega_z=1/4$ ): un-polarized, i.e, dipole moment set to zero 
(dashed line), polarized (solid line), and ideal gas (dotted line). The total number of $^{52}$ Cr atoms is 100000.
\label{fig:pancake1} }
\end{figure}

In Fig.~(\ref{fig:pancake1}) we show the condensate fraction in the
pancake trap as a function of temperature. For comparison, we have
included results of an un-polarized gas by setting the dipole moment
$d=0$. It is seen that the effect of the polarization of the gas is to
decrease the condensate fraction at any given temperature, compared to
the non-polarized gas. As a result the critical temperature is also
reduced. This effect may be expected since in a pancake trap the
average dipolar interaction is repulsive, thus the thermal effect due
to polarization is similar to that of increasing the scattering
length.



\begin{figure}
\resizebox{3.5in}{!}{\includegraphics{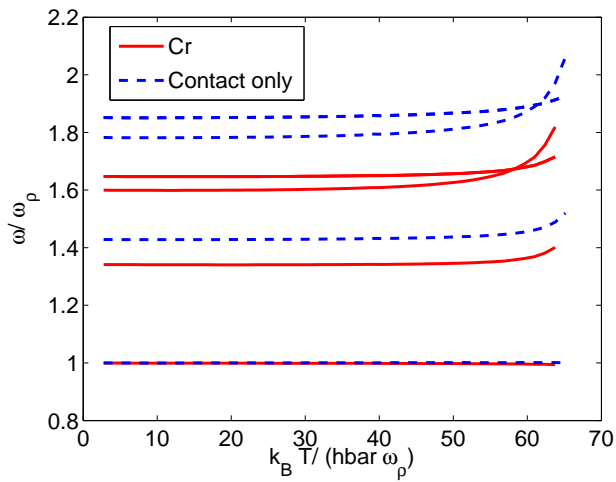}}
\caption{Lowest excitation frequencies a function of temperature for a  $^52$Cr condensate in a pancake trap with aspect ratio 1:4;  polarized (solid lines)and un-polarized (dashed lines). For each angular momentum number $m=0,1,2,3$, we plot only the lowest mode with
this $m$.
\label{fig:pancake3}}
\end{figure}

In Fig.~(\ref{fig:pancake3}) we plot the eigenfrequencies of the
lowest collective modes.Although the dipolar interaction causes large
shifts in the frequencies, the temperature dependence of these shifts
is very small, except very near the critical temperature. The shift in
frequencies due to polarization is, qualitatively, similar to that of
increasing the effective short range repulsion. Note the Kohn mode
($\omega/ \omega_z=1$) which should remain constant at 1. The slight
deviation from 1 at higher temperatures is due to Popov approximation
which in effect computes the dynamics of the condensate in the
presence of a static thermal component. A more correct description
should treat both components dynamically. Nevertheless, the deviation
of the Kohn mode from the theoretical value of 1 is small.

\subsection{Cr in a cigar trap}

\begin{figure}
\resizebox{3.5in}{!}{\includegraphics{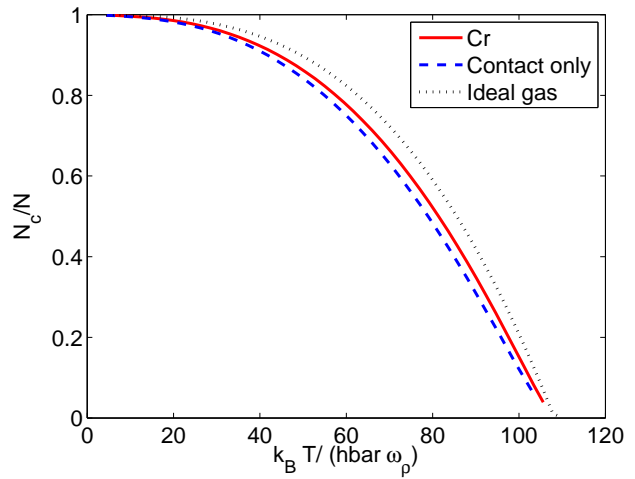}}
\caption{Condensate fraction as a function of temperature in a cigar
trap with aspect ratio 4:1; un-polarized (dashed line) polarized
(solid line), and ideal gas (dotted line).
\label{fig:cigar1}}
\end{figure}


\begin{figure}
\resizebox{3.5in}{!}{\includegraphics{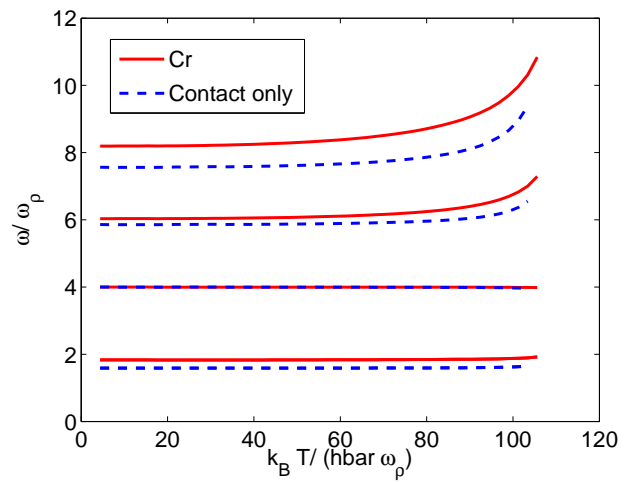}}
\caption{Lowest excitation frequencies as a function of temperature
for a $^52$Cr condensate in a cigar trap with aspect ratio 4:1;
polarized (solid lines)and un=polarized (dashed lines). Only modes
which are even in the $z$ direction are shown.For each angular
momentum number $m=0,1,2,3$, we plot only the lowest even mode with
this $m$.
\label{fig:cigar3}}
\end{figure}

Figs.~(\ref{fig:cigar1}) and (\ref{fig:cigar3})
show the condensate fraction and lowest collective
mode frequencies as a function of temperature for a $^{52}Cr$ gas in a
cigar trap with $\omega_z=2\pi \times 100$ Hz and $\omega_{\rho}=2\pi
\times 400$ Hz (with a reduced scattering length $a=20a_0$ as
before). Notice that now the effect of the dipolar interaction is to
increase the condensate fraction at any given temperature, thus
increasing also the critical temperature for the onset of
condensation. Again, this can be understood due to the dipolar
interaction being effectively attractive in a cigar
geometry. Similar to the the case of a pancake
trap, the dipolar interaction leads to significant shift in the
frequencies of the low modes, but these shifts depend only weakly on
temperature.

\subsection{Bi-Concave condensates}

We now turn to examine the finite temperate effects on the bi-concave
shaped condensate reported in Ref.~\cite{Ronen07}. There, we found an
interesting novel structure of pure dipolar condensates in pancake
traps at zero temperature. For appropriate choice of parameters, the
condensate density does not obtain its maximum in the center of the
trap. Rather, the maximum density is obtained along a ring, and the
center of the trap is local minimum of the density. This gives rise to
a bi-concave condensate shape similar to that of a red-blood
cell.Recently, other shapes have been predicted in non-cylindrically
symmetric traps \cite{Dutta07}. In this section, we investigate the
temperature effect on the bi-concave condensate in a cylindrically
symmetric trap.

\begin{figure}
\resizebox{3.5in}{!}{\includegraphics{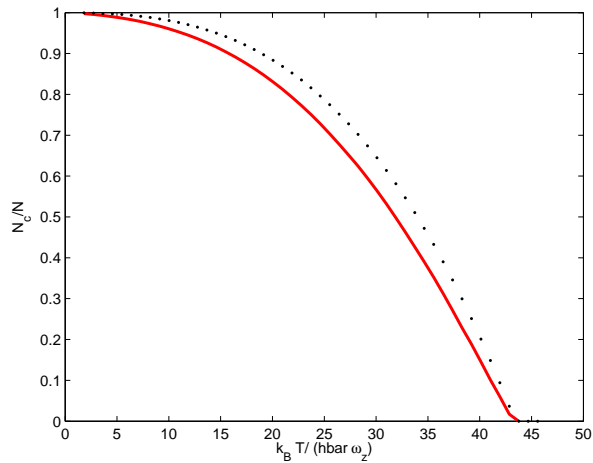}}
\caption{Condensate fraction as a function of temperature in a pancake
trap with aspect ratio 1:7, for which bi-concave structure is formed at
T=0; un-polarized (dashed line) polarized (solid line), and ideal gas
(dotted line). The total number of $^{52}$Cr atoms is 16300. The scattering length is 0.
\label{fig:cookie1}}
\end{figure}

In Fig.~(\ref{fig:cookie1}) we plot the condensate fraction as a
function of temperature for a pure $^{52}$Cr dipolar condensate (i.e,
where the scattering length has been tuned to zero via a Feshbach
resonance), in a pancake trap with aspect ratio
$\omega_{\rho}/\omega_z=1/7$. For number of particles $N=16300$, a
bi-concave structure is formed at $T=0$. In this figure it is notable
that the dipolar interaction brings about significant change (about
10\%) in the condensate fraction for tempeartures of order half the
critical temperature. Yet, the critical temperature itself is almost
un-changed. Indeed, the analytical formula of
Refs.~\cite{Glaum2007a,Glaum2007b} predicts a very small
reduction of 0.5\% in the critical temperature.

\begin{figure}
\resizebox{3.5in}{!}{\includegraphics{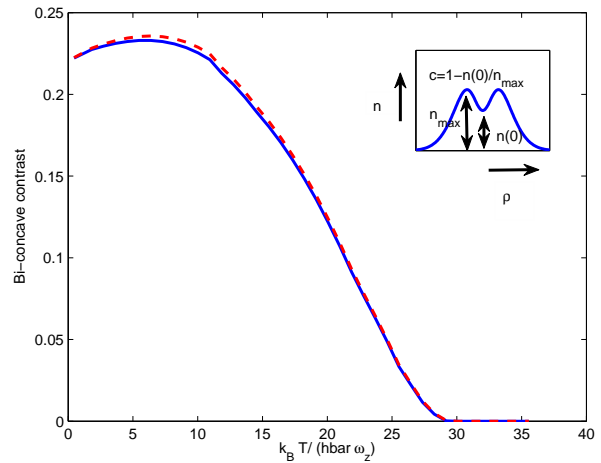}}
\caption{Biconcave contrast as a function of temperature in a pancake
trap with aspect ratio 1:7. Solid line: contrast parameter for the
total density; Dashed line: contrast parameter for the condensed part. Inset: illustration of
a typical biconcave density profile showing how the biconcave contrast $c$ is defined.
\label{fig:cookie2}}
\end{figure}

An interesting question is: what happens to the biconcave structure
with the increase in temperature? The Bi-concave shape eventually
washes out by $T=T_c$, yet (as we shall show) the shape persists to
surprisingly high temperature. To study this, We define a contrast
parameter, $c=1-n(0)/n_{\max}$, where $n(0)$ is the central density and
$n_{max}$ is the maximal density. For a normal
density profile where the maximal density is obtained at the center,
$c=0$.  In Fig.~(\ref{fig:cookie2}) we plot the biconcave structure
parameter for the total density profile, as well as for the condensed
part alone, as a function of temperature. It is seen that when the
temperature approaches about 70\% of the critical temperature, the
biconcave structure disappears. Generally, one would expect the
disappearance of the biconcave structure due to the thermal
excitations. We note, that according to Ref. \cite{Ronen07}, the
biconcave contrast (at $T=0$) is reduced with decreased number of
particles. Thus, when the condensate is depleted, we also expect the
bi-concave parameter to decrease. For $T>0.15 T_c$, this is indeed the
case. But for lower temperatures, we see that the biconcave contrast,
for both the total density and the condensed part alone, slightly
increases with temperature.

To understand this effect, let us first consider the density profile
of the thermal cloud alone. Consider first the simplest case of an
ideal gas in a harmonic trap. The thermal cloud occupies harmonic
oscillator states according to Bose statistics. The lowest and most
populated excited state, one above the ground state, has a node at
the center of the trap. Thus, at low temperatures, one expects the
thermal cloud to have reduced density at the center of the trap, even
in the absence of repulsive short range interactions. This effect is
easily verified by numerical simulations, and is seen to be more
pronounced when the dimensionality is reduced, such as in highly
pancake or cigar traps.  Of course, for an ideal gas, the total density
(thermal+condensate) still has its maximum at the center.

\begin{figure}
\resizebox{3.5in}{!}{\includegraphics{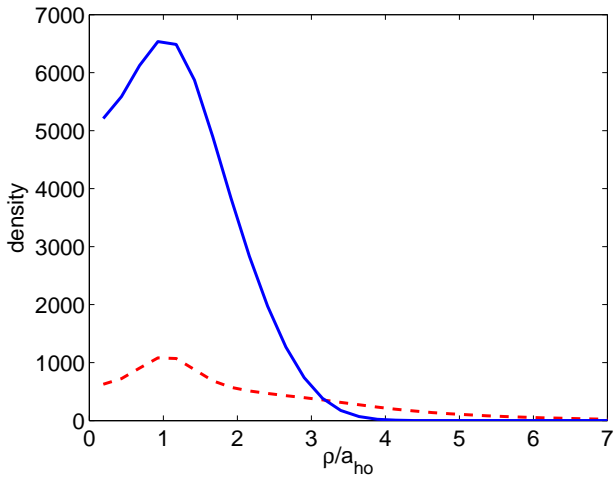}}
\caption{Radial density profiles of the condensate density (solid
line) and 10x the thermal component density (dashed line). Note that
we scaled the thermal component density by a factor of 10 for visual
comparison.  The trap aspect ratio is 1:7 as in
Fig.~\ref{fig:cookie2}, and the temperature is $T=0.2T_c$, where $T_c$
is the critical temperature. 
\label{fig:profiles}}
\end{figure}

\begin{figure}
\resizebox{3.5in}{!}{\includegraphics{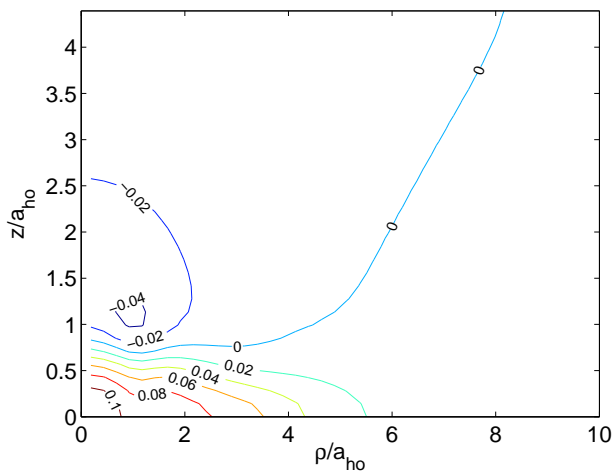}}
\caption{The mean dipolar field due to the thermal component density whose
radial profile is shown at Fig.~\ref{fig:profiles}.  
\label{fig:meanfield}}
\end{figure}

Consider now the dipolar gas. As the temperature is raised from $T=0$, the
lowest and most populated thermal gas mode is a sloshing (Kohn mode)
with a node in the center.  Thus for low temperatures the thermal
component has lower density at the center, as is demonstrated in
Fig.~(\ref{fig:profiles}).  The thermal component creates in turn a mean
field, which is shown in Fig.~(\ref{fig:meanfield}). It shows that the
maximum mean field potential is obtained in the center of the trap,
even though this is not where the place of maximum thermal component density. This is
due to the long range nature of the dipolar interaction: the
contributions from the ring of maximal density of the thermal cloud
add together in the center of trap. The mean field due to the thermal
cloud causes the condensate part to be repelled from the
center. The result is that the biconcave contrast of the condensate
is larger than it would have been with the same number of condensed
particles in a pure harmonic trap with no thermal component. This gives rise to the behavior seen
in Fig.~(\ref{fig:cookie2}) at low temperatures.

A caveat is that, as mentioned above, we made an approximation in our
computations by ignoring thermal exchange effects. This may, in
principle modify the effect of the thermal component on the condensed
part, at low temperatures in particular.  Exploring the effect of the
thermal exchange interaction would require a considerably heavier
computation than undertaken in this current work.

\begin{figure}
\resizebox{3.5in}{!}{\includegraphics{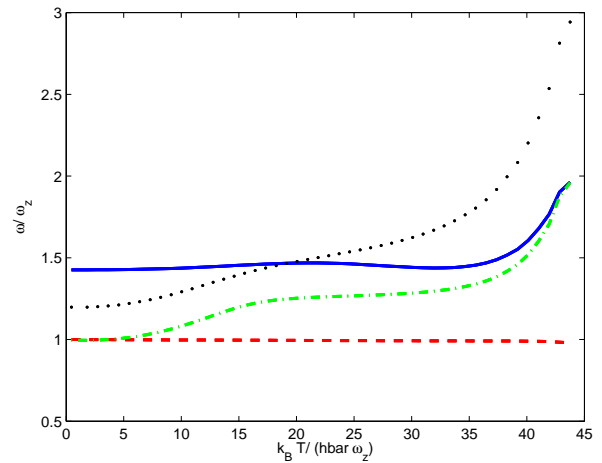}}
\caption{Lowest excitation frequencies as a function of temperature
for a $^52$Cr condensate in a pancake trap with aspect ratio
1:7. Solid line: lowest $m=0$ mode;dashed line: $m=1$ mode; dash dotted
line: $m=2$ mode; dots: $m=3$ mode;
\label{fig:cookie3}}
\end{figure}

In Fig.~(\ref{fig:cookie3}) we plot the lowest excitation
frequencies as a function of temperature for the pancake trap with
aspect ratio 1:7, (containing the above mentioned bi-concave
structure). The $m=1$ is as usual the Kohn mode. Note that at $T=0$
the $m=2$ and $m=3$ modes are lower than the $m=1$ mode. The near
degeneracy of the $m=1$ and $m=2$ modes at $T=0$ is accidental: for
higher number of particles, the $m=2$ and $m=3$ modes actually goes
below the $m=1$ mode, a consequence of the discrete roton-like
spectrum discussed in Ref. \cite{Ronen07}. Close to the critical
temperature the excitation energies approach their ideal gas values,
as might be expected due to the decreasing  density of the gas. In
between there is an interesting crossing between the $m=0$ and $m=3$
modes.

\subsection{Comparison with Monte Carlo}
Finally, we have attempted to compare the HFB-Popov method with the
path integral Monte Carlo simulations of Nho and Landau
\cite{Nho05}. However we do not find a good agreement. The energies
with HFB-Popov approximation for a dipolar condensate in a pancake
trap came about 10\% higher than the Monte Carlo simulation, with a
similar discrepancy in the shape (width) of the dipolar
condensates. We note that the Monte Carlo simulations were performed
for a very small number of particles, between 27 and 125. Under these
conditions,the critical temperature is very small, of the order of the
trap frequencies, and in fact lower than the chemical potential. Thus,
only a few low modes are excited even close to the critical
temperature $T_c$.  At the simulated temperature of $0.4 T_c$ the
density of the thermal gas is of the order of that of the condensate
and there is a large overlap between the two. Under these conditions
it may be expected that the approximation of ignoring the
thermal-thermal dipolar exchange interactions is invalid. Thus, it is
plausible that the disagreement is due to this additional
approximation rather than the inadequacy of the HFB-Popov method. But,
for a number of particles of order $10^4-10^5$, the critical
temperature is much higher than the trap frequencies and the density
of the thermal cloud significantly lower than that of the condensate.
Thus, we believe our method should still give valid results under
normal experimental conditions.

\section{Conclusions}

In conclusion, we applied the Hartree-Fock-Bogoliubov-Popov
approximation to dipolar gases in harmonic, cylindrically symmetric
traps. For computational reasons, the exchange interaction due to the
thermal gas has to be ignored (i.e, we ignore exchange interaction due
to long range spatial correlation in the thermal component). For
normal configurations where the condensate structure at $T=0$ has a
maximum at the center, we observe a temperature dependent behavior
similar to that of a gases with contact interaction
\cite{Hutchinson97,Dodd98}, but the behavior depends on the aspect
ratio of the trap: for pancake traps, the dipolar interaction is
effectively repulsive, leading to reduction of condensed part at a
given temperature, and thus also reduction of the critical temperature
for condensation, while in cigar trap, it is effectively attractive,
leading to the opposite thermal effects. For configurations where
bi-concave structure exists at $T=0$, we find the somewhat surprising
result that this structure is actually enhanced at low temperatures
(i.e, the ratio of the central density to the maximal density is
reduced). For higher temperatures, the bi-concave structure gradually
becomes less distinct and it disappears at a temperature of about 75\%
of the critical temperature. The low excitation spectrum of the
bi-concave structure also shows interesting temperature dependence
with crossing between different modes.

\begin{acknowledgments}
We gratefully acknowledges financial support from NSF.
\end{acknowledgments}

\bibliography{biblo}

\end{document}